\documentclass[11pt]{article}

\usepackage[preprint]{acl}

\usepackage{times}
\usepackage{latexsym}
\usepackage[T1]{fontenc}
\usepackage[utf8]{inputenc}
\usepackage{microtype}    
\usepackage{inconsolata}   
\usepackage{graphicx}     
\usepackage{float}

\usepackage{booktabs}    
\usepackage{amsmath}

\usepackage{xcolor} % needed for \todo above

\title{DevIntent: How Much Does LLM-Generated Code Violate Developer Intent?}

\author{Susana Haing \\
  Anote AI \\
  \texttt{shaing@ucsd.edu} \\\And
  Natan Vidra \\
  Anote AI \\
  \texttt{nvidra@anote.ai} \\\And
  Spurthi Setty \\
  Anote AI \\
  \texttt{ssetty2@stevens.edu}}
\begin{document}
\maketitle

% ============================================================
% ABSTRACT
% ============================================================
\begin{abstract}
Code generated by LLMs can violate a developer's implicit intentions when given an ambiguous prompt, yet standard benchmarks measure only whether code passes its stated test. We introduce the Intent Violation Rate (IVR) and a 49-problem pilot benchmark derived from HumanEval+. Each problem strips implicit constraints from a clarified prompt and encodes them as hidden constraint tests. IVR measures the fraction of LLM-generated solutions that pass the stated (visible) tests yet fail hidden constraint tests that capture unstated intent. Evaluating Claude Sonnet 4.6 and OpenAI GPT 4.1, we find both pass over 92\% of stated tests yet violate intent in over half of problems (54.5\% and 63.5\%), following a systematic, bimodal pattern consistent across both models. Out findings indicate that pass rates overstate how well generated code reflects developer intent. 
\end{abstract}

% ============================================================
% 1. INTRODUCTION 
% ============================================================
\section{Introduction}

When LLMs are tasked with building code, they can generate code sufficient for the main task at hand, such as writing a function or script. However, these outputs are not always fully correct, requiring human-in-the-loop review and adjustments before they can be deployed. Developers, through review, will capture functional or logical errors, and cases where the code does something different from what they intended. Current benchmarks measure functional correctness of LLM-generated code via a test pass rate \citep{austin2021program}, however, this metric does not capture whether the solution fully reflects what the developer intended. 

Test pass rates alone are insufficient as a metric because they do not detect intent violations, which refer to scenarios where LLMs will provide outputs that satisfy stated tests, but violate implicit constraints the developer assumes will be accounted for, and did not explicitly state. Prior work has documented similar failures in which LLMs produce code that will game specifications and exploit evaluation loopholes to pass tests without solving the intended task \citep{zhong2025impossiblebench}. We focus on a distinct failure mode: outputs that legitimately pass valid stated tests, but still violate unstated developer intent. To measure this, we propose the Intent Violation Rate (IVR), a metric that quantifies whether an output violates developer intent. We evaluate IVR on 49 algorithm specification problems, finding that across two models, they violate intent in more than half of the cases, with Sonnet 4.6 violating implicit developer constraints in approximately 54.5\% of cases, and GPT 4.1 in 63.5\%, even when solutions pass the stated tests.

% ============================================================
% 2. RELATED WORK
% ============================================================
\section{Related Work}
A key concern is when developers take the outputted code at face value due to a high pass rate on tests they built. Prior work defines code that passes provided test cases as having functional correctness \citep{austin2021program}, however \citet{chen2021codex} also note that models can pass weak test suites simply by deleting failing functionality. Models may also "overfit to assert statements", rather than properly apply logic \citep{austin2021program}, hardcoding outputs for visible tests rather than create an accurate function that properly handled the general task at hand. A growing body of work also details LLMs that exploit evaluation shortcuts in code generation; this is seen in test suites where tasks were constructed to purposefully contradict specifications \citep{zhong2025impossiblebench}, or where outputs are hardcoded, or tests were edited to pass \citep{gabor2025evilgenie}. Despite the focus on deliberate exploitation of flawed or gameable test suites, they further show that solutions passing held-out tests can still be incorrect. Recent benchmarks also expand evaluation beyond basic generation, such as LiveCodeBench \citep{jain2024livecodebench}, which adds tasks like self-repair and code execution while addressing contamination through time-segmented problem collection. These efforts nonetheless measure functional correctness, and we build on this observation to measure violations of unstated intent in generated code that passes valid stated tests. 

By default, models rarely ask for clarification when a prompt is unclear, so when it comes to giving LLMs ambiguous prompts, prior work has seen a drop in test pass rates by approximately 80\% \citep{li2026clareval}. It shows that functional correctness (Pass@1) will fall apart when ambiguity is introduced to prompts, and we measure the intent violated in code where functional correctness holds. However, to address the drop in test pass rates, other work finds that asking clarifying questions improves test pass rates \citep{mu2023clarifygpt}, and ambiguity-aware training improves detection \citep{kim-etal-2024-aligning}. Rather than utilize methods to resolve ambiguity through interaction or prompt repair, IVR measures intent violations that remain in code that passes its stated tests, generated from ambiguous prompts. 

% ============================================================
% 3. METHOD
% ============================================================
\section{Measuring Intent Violation Rate}

\subsection{Definition}
Intent Violation Rate, or IVR, is defined as follows: 
\begin{equation}
  \mathrm{IVR} = \frac{n_{\text{violating}}}{n_{\text{stated passed}}}
\end{equation}

Where for a given solution:
\begin{description}
  \item[$n_{\text{stated passed}}$:] number of generated solutions passing all stated tests
  \item[$n_{\text{violating}}$:] subset of those that fail $\geq 1$ hidden constraint test
\end{description}

IVR can be interpreted as: "Of the solutions that satisfy the given spec's visible tests, what fraction violates the developer's actual intent?" It is only measured over solutions that pass visible stated tests, therefore any solutions that do not pass the initial public test are not included in the calculation. The reported IVR is the mean of per-problem IVRs over problems that had $\geq 1$ solution passing stated test C1. 

\subsection{Benchmark Construction}
The built benchmark consists of 49 algorithm problems derived from HumanEval+ \citep{liu2023evalplus}. Each problem consists of an ambiguous prompt, a gold prompt, a stated test (C1), and hidden constraint tests (C2--C4, the number varying per problem).

The selected HumanEval+ problems all contained a gold prompt, which we stripped down to produce an ambiguous prompt. In some cases, additional specifics were added to the gold prompt to further clarify the task at hand. This gold prompt was always hidden to the LLM, as it only received the ambiguous prompt, and C1, the stated test. Each aspect that was stripped from the gold prompt to formulate the ambiguous prompt was then used to create the hidden constraint tests. 

Ambiguous prompts were constructed manually by the authors, removing explicitly stated constraints from each gold prompt while preserving the core task description. For a subset of problems, a Claude LLM was used as a drafting aid for the C2-C4 hidden constraint tests in cases where it was unclear whether a hand-written test adequately captured the intended constraint. In all cases, the authors manually reviewed and validated the final tests, and all tests pass the validation checks described in Section 3.2

The scope was narrowed down to algorithm problems, where the prompt described a self contained function. A subset of these problems were cleanly decomposable into executable tests, and the formulation of the ambiguous prompt came from under specifying rather than misleading. Other task types, such as data transformation, involve fuzzier constraints and lack a comparable source at scale, being left to future work. 

As an example, the original HumanEval/34 prompt reads: \textit{"Return sorted unique elements in a list."}

We decompose it as shown in Table~\ref{tab:spec-example}. 
\begin{table}[ht]
\small
\centering
\begin{tabular}{@{}p{0.9cm}p{6cm}@{}}
\toprule
\multicolumn{2}{@{}l}{\textbf{HumanEval/34}} \\
\midrule
Ambig. & Return the unique elements in a list. \\
\addlinespace
Gold & Return only the unique values, sorted in
        ascending order. \\
\midrule
\textbf{C1} & \emph{stated}: only unique values \newline
  \texttt{solution([-2,4,4,6,6]) == [-2,4,6]} \\
\addlinespace
\textbf{C2} & \emph{hidden}: sorted ascending \newline
  \texttt{solution([-2,0,5,3,5,0,3])} \newline
  \texttt{== [-2,0,3,5]} \\
\bottomrule
\end{tabular}
\caption{Example spec pair. Stripping the sort-order
requirement from the gold prompt yields the ambiguous
prompt; C1 tests what remains stated, C2 tests the
stripped intent.}
\label{tab:spec-example}
\end{table}

To ensure the benchmark is sound, we applied two validation checks. First, a leakage audit was performed. When building the stated test, it must be satisfiable from the ambiguous prompt alone. During initial construction, 11 of the 49 problems failed this check. 9 of the failed tests were due to type errors, output format contamination, and hidden constraints leaking into the stated test. Of the hidden constraint leakage, what occurred was that the stated test also check a hidden constraint test, so a solution could not pass without already knowing part of the developer's intent. The other 2 problems were identified in the second validation check, where the canonical solution was run against the entire test suite to ensure stated and hidden tests were sound. These issues were errors in the tests themselves, something the initial leakage audit could not catch. Across all 49 problems, the validated references yield an IVR = 0, confirming that the hidden tests are satisfiable by correct code, and are not themselves the source of the measured violations. 

\subsection{Evaluation Pipeline}
The pipeline occurs in 4 stages: generation, execution, pass/fail determination, and IVR calculation. 

Generation: Solutions are generated from the ambiguous prompt and the stated test C1, using Claude Sonnet 4.6 and GPT 4.1, both at the API default temperature of 1.0. All other information (e.g., gold clarified prompt and hidden tests) is not shown to the LLM. 

Execution: For each problem, 5 solutions are produced, with a 5-second timeout per test, returning a pass/fail result. This is adapted from HumanEval's own execution pipeline, which runs candidate solutions in process isolation with syscall lockdown and timeouts \citep{chen2021codex}. 

Pass/Fail Determination: Each solution for each problem is run against the stated test C1 and hidden tests (C2--C4), producing a per-solution result for each problem. IVR is then calculated for that specific problem. 

Overall IVR Calculation: Each problem produces its own IVR value, and the overall IVR value is the mean across N problems with at least one stated-test-passing solution. 

95\% confidence intervals are also reported, estimated via bootstrap resampling over problems (10,000 iterations, seeded).

% ============================================================
% 4. EXPERIMENTS
% ============================================================
\section{Experiments}

\subsection{Setup}
We evaluated Sonnet 4.6 and GPT-4.1 following the pipeline in Section 3.3, generating 5 solutions per problem at temperature 1.0 with a 1024-token limit. This is a dual-model evaluation; we leave additional models for future work. 

\subsection{Results}
Across the stated tests, Sonnet 4.6 passes 94.3\% of the time, but violates at least one hidden intent constraint in 54.5\% of problems (95\% CI: 40.4-68.1\%); GPT-4.1 passes 92.7\% but violates in 63.5\% (95\% CI: 50.0-76.5\%). Both models exhibit the high functional correctness of LLM-generated code that nonetheless violates unstated intent. Of the 49 benchmark problems, some between the two models (Sonnet 4.6: HE/101, HE/112; GPT-4.1: HE/21, HE/108, HE/112) had no solution passing the stated test C1 and are therefore excluded from the IVR calculation per model, leaving N=47 for Sonnet 4.6 and N=46 for GPT-4.1. The confidence intervals overlap, so the two model's violation rates are not statistically distinguishable at this sample size, though substantial. 

\begin{table}[ht]
\centering
\begin{tabular}{@{}lcc@{}}
\toprule
\textbf{Metric} & \textbf{Claude Sonnet 4.6} & \textbf{GPT-4.1} \\
\midrule
C1 pass rate        & 94.3\% & 92.7\% \\
IVR                 & 54.5\% & 63.5\% \\
95\% CI  & 40.4--68.1\% & 50.0--76.5\% \\
Qualifying Problems & 47     & 46     \\
\% at IVR poles     & 95.7\% & 91.3\% \\
\bottomrule
\end{tabular}
\caption{IVR results for both models across the 49 spec
pairs (5 solutions per problem). Qualifying problems have $\geq 1$
stated-test-passing solution (excluded: HE/101, HE/112
for Claude; HE/21, HE/108, HE/112 for GPT-4.1). Per-problem
IVR concentrates at the extremes (0 or 1) for both models.}
\label{tab:summary-results}
\end{table}

\begin{figure}[ht]
\centering
\includegraphics[width=\columnwidth]{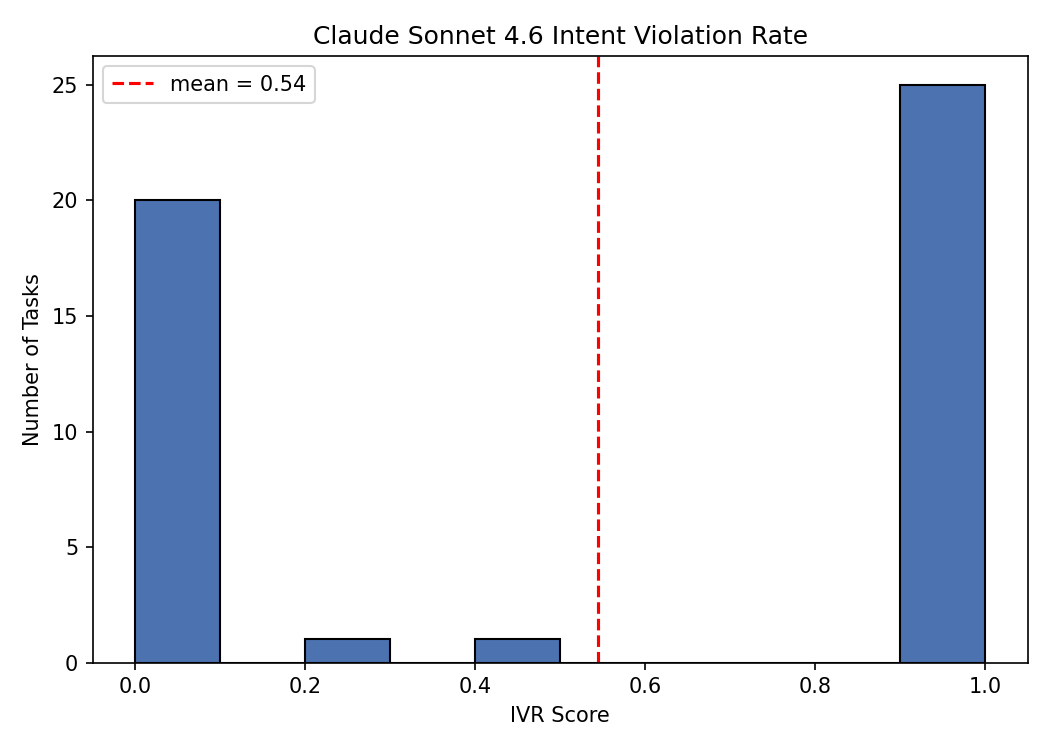}
\caption{Distribution of per-problem IVR for Claude
Sonnet 4.6 across 47 spec pairs. Two problems with no stated-test-passing solutions are excluded.}
\label{fig:claude_distribution}
\end{figure}

\begin{figure}[ht]
\centering
\includegraphics[width=\columnwidth]{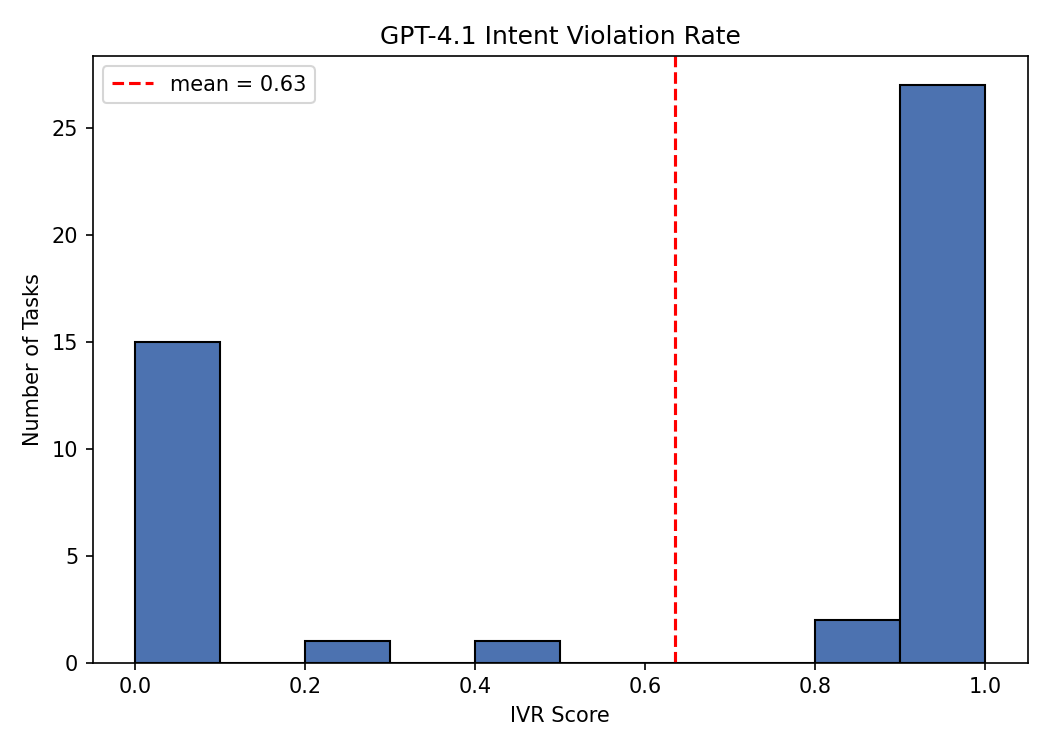}
\caption{Distribution of per-problem IVR for GPT 4.1 across 46 spec pairs. Three problems with no stated-test-passing solutions are excluded.}
\label{fig:OpenAI_distribution}
\end{figure}

Problems tended to cluster at IVR = 0 (no solutions violate intent) or IVR = 1 (every solution violates intent), with 45 of 47 (95.7\%) for Sonnet 4.6 (Figure~\ref{fig:claude_distribution}) and 42 of 46 (91.3\%) for GPT-4.1 (Figure~\ref{fig:OpenAI_distribution}) at these extremes, showing a bimodal distribution for both models. This near-deterministic behavior indicates intent violations are systematic, not sampling noise. If violations were due to random variation in generation, per-problem IVR would spread across intermediate values. Instead, both independently trained models produce outputs that either fail to or fully capture hidden constraints.

Both models fail stated test C1 much less often than the hidden constraints, at 5.7\% for Sonnet 4.6 and 7.3\% for GPT-4.1. Among the hidden constraints, fail rates decline monotonically from C2 to C4 for each model (Table~\ref{tab:constraint-breakdown}): from 42.9\% to 15.4\% for Sonnet 4.6 and from 44.1\% to 21.5\% for GPT-4.1. GPT-4.1's fail rates are somewhat higher, however, the overall difference is not significant. 

\begin{table}[ht]
\centering
\begin{tabular}{@{}lccc@{}}
\toprule
\textbf{Constraint} & \textbf{Present} & \textbf{Failed} & \textbf{Fail Rate} \\
\midrule
\multicolumn{4}{@{}l}{\textit{Claude Sonnet 4.6}} \\
C1 (stated) & 245 & 14 & 5.7\%  \\
C2 (hidden) & 231 & 99 & 42.9\% \\
C3 (hidden) & 201 & 58 & 28.9\% \\
C4 (hidden) & 65  & 10 & 15.4\% \\
\midrule
\multicolumn{4}{@{}l}{\textit{GPT-4.1}} \\
C1 (stated) & 245 & 18 & 7.3\%  \\
C2 (hidden) & 227 & 100 & 44.1\% \\
C3 (hidden) & 199 & 69 & 34.7\% \\
C4 (hidden) & 65  & 14 & 21.5\% \\
\bottomrule
\end{tabular}
\caption{Failure rate by constraint for both models across
the 49 spec pairs (245 solutions each). ``Present'' counts solutions whose problem
included the constraint; C2--C4 rates are over
stated-test-passing solutions.}
\label{tab:constraint-breakdown}
\end{table}

\subsection{Analysis}
Hidden constraints are numbered in the order they were decomposed from the gold prompt, and this ordering is positional and does not strictly rank constraints by centrality. We nonetheless observe that earlier numbered constraints fail more often than later ones for both models (Table~\ref{tab:constraint-breakdown}), which may indicate that the first stripped constraint is often most central to the task.

In HumanEval/34 (Section 3.2), the IVR of 1.0 for both models, with every stated-test-passing solution failing hidden ascending-sort order constraint C2. However, in HumanEval/25, the ambiguous prompt reads as "Write a function named solution that returns the prime factors of a given integer", and it contains the same hidden constraint of returning the output in ascending order. Across both models, all five solutions pass the stated and hidden constraints, yielding an IVR of 0.0, as standard factorization naturally produces factors in ascending order. This suggests intent satisfaction is constraint and context-dependent. 

% ============================================================
% 5. CONCLUSION 
% ============================================================
\section{Conclusion}

We introduced the Intent Violation Rate (IVR), a metric that measures how often LLM-generated code passes its stated tests while violating unstated developer intent, and a small pilot benchmark of N=49 audited algorithm specification pairs derived from HumanEval+. Evaluating Claude Sonnet 4.6 and GPT 4.1 on this benchmark, despite high stated-test pass rates (94.3\% and 92.7\%), both violated intent in over half of the problems (54.5\% and 63.5\%). These violations fell into a systematic and near-deterministic pattern in both models rather than random noise, indicating that functional correctness metrics can overstate how well generated code reflects what a developer actually intended. Future work includes underspecification detection, expanding the benchmark, and extending IVR across additional models and task domains. 

% ============================================================
% LIMITATIONS 
% ============================================================
\section*{Limitations}

Our experiment and benchmarking were only tested on the Sonnet 4.6 and GPT-4.1 models, at a single temperature of 1.0 and max tokens of 1024. IVR could differ across other models, such as Llama or Qwen and different settings such as lower temperatures. Additionally, the benchmark is small, at only N=49, limiting statistical power, with constraint-type distributions being more suggestive than robust. The overlapping confidence intervals also mean we cannot distinguish the two models' violation rates. 

Per-problem IVR is computed over only 5 samples, so it takes on one of six discrete values. This coarse resolution limits how precisely we can characterize the distribution's shape, however the strong concentration as the extremes suggest clustering is not solely an artifact in Section 4.2. 

A Claude model was used to assist in drafting a subset of the constraint tests, though all tests were manually reviewed and validated by the authors and passed our validation checks (Section 3.2). Due to this, Claude Sonnet 4.6's measured IVR could be biased downward, toward passing tests it may have helped shape, making its reported rate a conservative estimate. GPT-4.1, which played no role in test construction, serves as an independent control, exhibiting a comparable and higher IVR on the same tests, indicating that the measured violations are not an artifact of test authorship. 

The benchmark only focuses on Python algorithmic problems as other languages and problem types like data transformation may yield different findings. The problem specification is constrained to single-annotator construction, therefore the defined behavior of clarified gold prompts and tests reflects the judgment of one person. This also affects the operationalized definition of "intent", as we do not measure "intent" in an absolute metric. Also, HumanEval+ is a public dataset, so it is possible that the models have seen the original problems in training prior to our experimentation. 

% ============================================================
% ETHICAL CONSIDERATIONS
% ============================================================
\section*{Ethical Considerations}
Our work highlights a failure mode with direct deployment implications: LLM-generated code can pass its stated tests while silently violating unstated developer intent. As LLM-assisted coding is increasingly integrated into software workflows, a high test pass rate may give developers false confidence in code that does not do what they intended. We frame IVR as a measurement that  makes this gap measurable, and we caution against treating functional-correctness metrics as sufficient evidence of correctness in deployment. 

Our benchmark is derived from HumanEval+ \citep{liu2023evalplus}, which is publicly available, so it is possible the evaluated model encountered the original problems during training. We discuss this contamination risk in the Limitations section. We introduce no new human-subjects data, and the benchmark contains only algorithmic programming problems with no personal, sensitive, or identifying information. 

Finally, our operational definition of intent reflects the judgment of a single annotator constructing the ambiguous and gold prompts and the test suites. Any measure of intent is inherently normative, and our benchmark encodes one reasonable interpretation of each problem rather than a ground truth. We therefore present IVR as a relative diagnostic, not an absolute measure of whether LLM-generated code is "correct".

% ============================================================
% ACKNOWLEDGMENTS — camera-ready ONLY. Leave commented out
% for the anonymous submission.
% ============================================================
% \section*{Acknowledgments}
% TODO after acceptance

% ============================================================
% REFERENCES — unlimited pages
% Include DOIs/URLs in every entry where possible.
% ============================================================
\bibliography{custom}

% ============================================================
% APPENDICES — after references, letter-numbered, free space,
% double-column format (default; don't switch to onecolumn
% unless a section is math-heavy). Must be supplemental only:
% the paper must stand alone without them.
% ============================================================
\appendix

\section{Spec Pair Construction Details}
\label{app:construction}
Construction Protocol: 
- Start from a HumanEval+ problem and its original prompt 
- Identify the implicit constraints the original problem specifies (ordering, edge-case handling, conditional behavior, etc.)
- Write a gold (clarified) prompt built off the original prompt that states all constraints explicitly; in some cases add specifics beyond the original to remove residual ambiguity 
- Strip the constraints from the gold prompt to produce the ambiguous prompt; omitting information that does not steer toward a wrong interpretation 
- Encode each stripped constraint as a hidden test (C2--C4); C1 is a stated test that must be satisfiable from the ambiguous prompt alone 
- Problems are algorithm only as they cleanly decompose into executable tests; other task types involve fuzzier constraints and lack a comparably licensed source at scale

\begin{table}[ht]
\small
\centering
\begin{tabular}{@{}p{0.9cm}p{6cm}@{}}
\toprule
\multicolumn{2}{@{}l}{\textbf{HumanEval/88} (conditional sort, IVR $=1.0$)} \\
\midrule
Ambig. & Return a sorted copy of a list of non-negative
         integers, without modifying the original. \\
\addlinespace
Gold & Sort ascending if the sum of the first and last
       elements is odd, descending if even; do not mutate
       the input. \\
\midrule
\textbf{C1} & \emph{stated}: sorts (odd sum) \newline
  \texttt{solution([2,4,3,0,1,5])} \newline
  \texttt{== [0,1,2,3,4,5]} \\
\addlinespace
\textbf{C2} & \emph{hidden}: descending on even sum \newline
  \texttt{solution([2,4,3,0,1,5,6])} \newline
  \texttt{== [6,5,4,3,2,1,0]} \\
\bottomrule
\end{tabular}
\caption{HE/88: the model sorts but ignores the
parity-dependent direction. All 5 solutions violate C2.}
\label{tab:ex-88}
\end{table}

\begin{table}[ht]
\small
\centering
\begin{tabular}{@{}p{0.9cm}p{6cm}@{}}
\toprule
\multicolumn{2}{@{}l}{\textbf{HumanEval/161} (fallback, IVR $=1.0$)} \\
\midrule
Ambig. & Swap the case of each letter, leaving non-letter
         characters unchanged. \\
\addlinespace
Gold & Swap the case of each letter; if the string
       contains no letters, return it reversed instead. \\
\midrule
\textbf{C1} & \emph{stated}: case swap \newline
  \texttt{solution('\#a@C') == '\#A@c'} \\
\addlinespace
\textbf{C2} & \emph{hidden}: reverse if no letters \newline
  \texttt{solution('1234') == '4321'} \\
\bottomrule
\end{tabular}
\caption{HE/161: the model performs the case swap but
misses the no-letters fallback. All 5 solutions violate C2.}
\label{tab:ex-161}
\end{table}

\begin{table}[ht]
\small
\centering
\begin{tabular}{@{}p{0.9cm}p{6cm}@{}}
\toprule
\multicolumn{2}{@{}l}{\textbf{HumanEval/25} (satisfied, IVR $=0.0$)} \\
\midrule
Ambig. & Return the prime factors of a given integer. \\
\addlinespace
Gold & Return the prime factors in ascending order, with
       repeated factors and product equal to the input. \\
\midrule
\textbf{C1} & \emph{stated}: basic factorization \newline
  \texttt{solution(13) == [13]} \\
\addlinespace
\textbf{C2} & \emph{hidden}: ascending order \newline
  \texttt{solution(15) == [3,5]} \\
\bottomrule
\end{tabular}
\caption{HE/25: the ascending-order constraint is
satisfied by all 5 solutions, as standard factorization
emits factors in ascending order. Contrast with HE/34,
where the analogous ordering constraint is violated.}
\label{tab:ex-25}
\end{table}

\subsection*{Validation Detail}
The benchmark was validated in two passes.

\textbf{Leakage audit.} Each stated test (C1) must be
satisfiable from the ambiguous prompt alone, without
knowledge of any hidden constraint. Of the 49 problems,
9 initially failed this check: type errors, output-format
contamination (e.g., a list expected where a tuple was
returned), and hidden-constraint leakage, where C1
inadvertently tested a hidden constraint so that no
solution could pass C1 without already encoding part of
the intent. All 9 were corrected.

\textbf{Canonical soundness check.} We ran the HumanEval+
canonical solution for each problem against the full test
suite (C1 and all hidden constraints); a correct solution
must pass all constraints, yielding IVR $=0$. This surfaced
2 further problems whose tests were themselves faulty
rather than leaked: HE/47 had an incorrect expected median
value, and HE/18 contained a contradictory empty-substring
constraint. Both were corrected (the HE/18 constraint was
removed, as the empty-substring convention is ambiguous).
After correction, all 49 canonical references yield
IVR $=0$, confirming the hidden tests are satisfiable by
correct code.

\section{Prompts and Sandbox Configuration}
\label{app:prompts}
The generation prompt template is:
\begin{verbatim}
Write a Python function named `solution` that 
solves the following task.
{prompt}
Requirements:
- Name the function exactly `solution`.
- Return only the function definition — no 
  explanation, no markdown, no imports 
  unless the function itself needs them.
\end{verbatim}
where \texttt{\{prompt\}} is the ambiguous prompt.

Decoding Parameters: 
- Models: Claude Sonnet 4.6 and OpenAI GPT-4.1
- Temperature: 1.0 (API default)
- Max Tokens: 1024 
- Number Samples per Problem: 5 

Post Processing: Markdown code fences are stripped from model output before execution to prevent syntax errors from non-code formatting.

Sandbox: Subprocess-based execution, one subprocess per test, 5-second timeout per test, adapted from HumanEval's execution harness (process isolation, syscall restrictions). Each solution is run against C1 (stated test) and each hidden constraint independently. 

\section{Claude Sonnet 4.6 Per-Problem Results}
\label{app:results}
Table~\ref{tab:per-problem-claude} reports the per-problem IVR for all 49 spec pairs,
the number of solutions passing the stated test (C1) out of 5, and which hidden
constraints were violated. HE/101 and HE/112 had no C1-passing solutions and are
excluded from the IVR mean.

\begin{table}[ht]
\centering
\small
\begin{tabular}{@{}lccl@{}}
\toprule
\textbf{Problem} & \textbf{C1 pass} & \textbf{IVR} & \textbf{Violated} \\
\midrule
HE/7 & 5/5 & 0.00 & -- \\
HE/12 & 5/5 & 0.40 & C3 \\
HE/18 & 5/5 & 0.00 & -- \\
HE/20 & 5/5 & 0.00 & -- \\
HE/21 & 5/5 & 0.00 & -- \\
HE/25 & 5/5 & 0.00 & -- \\
HE/26 & 5/5 & 0.00 & -- \\
HE/30 & 5/5 & 0.00 & -- \\
HE/34 & 5/5 & 1.00 & C2 \\
HE/42 & 5/5 & 0.00 & -- \\
HE/47 & 5/5 & 0.00 & -- \\
HE/51 & 5/5 & 0.00 & -- \\
HE/57 & 5/5 & 1.00 & C2 \\
HE/58 & 5/5 & 1.00 & C2,C3 \\
HE/64 & 5/5 & 1.00 & C2,C3 \\
HE/65 & 5/5 & 1.00 & C3 \\
HE/66 & 5/5 & 1.00 & C2 \\
HE/68 & 5/5 & 0.00 & -- \\
HE/70 & 5/5 & 0.00 & -- \\
HE/71 & 5/5 & 1.00 & C2,C3 \\
HE/72 & 5/5 & 1.00 & C2 \\
HE/74 & 5/5 & 0.00 & -- \\
HE/86 & 5/5 & 1.00 & C2,C3 \\
HE/88 & 5/5 & 1.00 & C2 \\
HE/90 & 5/5 & 1.00 & C3,C4 \\
HE/91 & 5/5 & 1.00 & C2,C3 \\
HE/93 & 5/5 & 1.00 & C2,C3 \\
HE/95 & 5/5 & 1.00 & C4 \\
HE/98 & 5/5 & 0.00 & -- \\
HE/99 & 5/5 & 0.00 & -- \\
HE/101 & 0/5 & -- & -- \\
HE/102 & 5/5 & 1.00 & C3 \\
HE/104 & 5/5 & 1.00 & C2 \\
HE/105 & 5/5 & 0.20 & C2 \\
HE/108 & 1/5 & 0.00 & -- \\
HE/111 & 5/5 & 1.00 & C2,C3 \\
HE/112 & 0/5 & -- & -- \\
HE/120 & 5/5 & 0.00 & -- \\
HE/126 & 5/5 & 1.00 & C3 \\
HE/131 & 5/5 & 0.00 & -- \\
HE/135 & 5/5 & 0.00 & -- \\
HE/136 & 5/5 & 0.00 & -- \\
HE/140 & 5/5 & 1.00 & C2 \\
HE/145 & 5/5 & 1.00 & C2 \\
HE/149 & 5/5 & 1.00 & C2 \\
HE/151 & 5/5 & 1.00 & C2 \\
HE/158 & 5/5 & 1.00 & C2 \\
HE/161 & 5/5 & 1.00 & C2 \\
HE/163 & 5/5 & 1.00 & C2,C3 \\
\bottomrule
\end{tabular}
\caption{Per-problem IVR across all 49 spec pairs. ``C1 pass'' is the
number of the 5 sampled solutions passing the stated test; ``IVR'' is the
fraction of those that violate $\geq 1$ hidden constraint; ``Violated''
lists the hidden constraints failed by any C1-passing solution.}
\label{tab:per-problem-claude}
\end{table}

\section{OpenAI GPT-4.1 Per-Problem Results}
\label{app:results-gpt}
Table~\ref{tab:per-problem-gpt} reports the per-problem IVR for GPT-4.1 across
all 49 spec pairs. HE/21,
HE/108, and HE/112 had no C1-passing solutions and are excluded from the IVR mean.

\begin{table}[ht]
\centering
\small
\begin{tabular}{@{}lccl@{}}
\toprule
\textbf{Problem} & \textbf{C1 pass} & \textbf{IVR} & \textbf{Violated} \\
\midrule
HE/7 & 5/5 & 0.00 & -- \\
HE/12 & 5/5 & 1.00 & C3 \\
HE/18 & 5/5 & 0.80 & C2 \\
HE/20 & 5/5 & 0.00 & -- \\
HE/21 & 0/5 & -- & -- \\
HE/25 & 5/5 & 0.00 & -- \\
HE/26 & 5/5 & 0.00 & -- \\
HE/30 & 5/5 & 0.00 & -- \\
HE/34 & 3/5 & 1.00 & C2 \\
HE/42 & 5/5 & 0.00 & -- \\
HE/47 & 5/5 & 0.00 & -- \\
HE/51 & 5/5 & 0.00 & -- \\
HE/57 & 5/5 & 1.00 & C2 \\
HE/58 & 5/5 & 1.00 & C3 \\
HE/64 & 5/5 & 1.00 & C2,C3 \\
HE/65 & 5/5 & 1.00 & C3 \\
HE/66 & 5/5 & 1.00 & C2 \\
HE/68 & 5/5 & 0.80 & C3,C4 \\
HE/70 & 5/5 & 0.00 & -- \\
HE/71 & 5/5 & 1.00 & C2,C3 \\
HE/72 & 5/5 & 1.00 & C2 \\
HE/74 & 5/5 & 0.40 & C2 \\
HE/86 & 5/5 & 1.00 & C2,C3 \\
HE/88 & 5/5 & 1.00 & C2 \\
HE/90 & 5/5 & 1.00 & C3,C4 \\
HE/91 & 5/5 & 1.00 & C2,C3 \\
HE/93 & 5/5 & 1.00 & C2,C3 \\
HE/95 & 5/5 & 1.00 & C4 \\
HE/98 & 5/5 & 0.00 & -- \\
HE/99 & 5/5 & 0.00 & -- \\
HE/101 & 4/5 & 1.00 & C2,C3 \\
HE/102 & 5/5 & 1.00 & C3 \\
HE/104 & 5/5 & 1.00 & C2 \\
HE/105 & 5/5 & 0.20 & C2 \\
HE/108 & 0/5 & -- & -- \\
HE/111 & 5/5 & 1.00 & C2 \\
HE/112 & 0/5 & -- & -- \\
HE/120 & 5/5 & 0.00 & -- \\
HE/126 & 5/5 & 1.00 & C3 \\
HE/131 & 5/5 & 0.00 & -- \\
HE/135 & 5/5 & 0.00 & -- \\
HE/136 & 5/5 & 0.00 & -- \\
HE/140 & 5/5 & 1.00 & C2,C3 \\
HE/145 & 5/5 & 1.00 & C2 \\
HE/149 & 5/5 & 1.00 & C2 \\
HE/151 & 5/5 & 1.00 & C2 \\
HE/158 & 5/5 & 1.00 & C2 \\
HE/161 & 5/5 & 1.00 & C2 \\
HE/163 & 5/5 & 1.00 & C2,C3 \\
\bottomrule
\end{tabular}
\caption{Per-problem IVR for GPT-4.1 across all 49 spec
pairs. ``C1 pass'' is the number of the 5 sampled solutions
passing the stated test; ``IVR'' is the fraction of those
that violate $\geq 1$ hidden constraint; ``Violated'' lists
the hidden constraints failed by any C1-passing solution.}
\label{tab:per-problem-gpt}
\end{table}

\end{document}